\documentclass[conference]{IEEEtran}
\IEEEoverridecommandlockouts

\usepackage{cite}
\usepackage{hyperref}
\usepackage{subcaption}

\hypersetup{
    colorlinks=true,     
    linkcolor=blue,       
    citecolor=blue,     
    urlcolor=blue,       
    }

\usepackage{amsmath,amssymb,amsfonts}
\usepackage{algorithmic}
\usepackage{graphicx}
\usepackage{textcomp}
\usepackage{xcolor}
\usepackage{enumitem}
\makeatletter
\def\@cite#1#2{\textcolor{blue}{[{#1\if@tempswa , #2\fi}]}}
\makeatother
\def\BibTeX{{\rm B\kern-.05em{\sc i\kern-.025em b}\kern-.08em
    T\kern-.1667em\lower.7ex\hbox{E}\kern-.125emX}}
\begin{document}

\title{SpectrumFM: Redefining Spectrum Cognition via Foundation Modeling}

\author{\IEEEauthorblockN{Chunyu Liu$^{\S}$, Hao Zhang$^{\S}$, Wei Wu$^{\ddagger}$, Fuhui Zhou$^{\S}$, Qihui Wu$^{\S}$, \\Derrick Wing Kwan Ng$^{\ast }$, and Chan-Byoung Chae$^{\star }$ \\
$^{\S}$Nanjing University of Aeronautics and Astronautics, Nanjing, China, \\
$^{\ddagger}$Nanjing University of Posts and Telecommunications, Nanjing, China, \\
$^{\ast }$ The University of New South Wales, Sydney, Australia,
$^{\star }$ Yonsei University, Seoul, South Korea.\\
Email: \emph{chunyu.liu@nuaa.edu.cn, haozhangcn@nuaa.edu.cn, weiwu@njupt.edu.cn,} \\
\emph{zhoufuhui@ieee.org, wuqihui2014@sina.com, w.k.ng@unsw.edu.au, cbchae@yonsei.ac.kr} 
}
}
\maketitle

\begin{abstract}
The enhancement of spectrum efficiency and the realization of secure spectrum utilization are critically dependent on spectrum cognition. However, existing spectrum cognition methods often exhibit limited generalization and suboptimal accuracy when deployed across diverse spectrum environments and tasks. 
To overcome these challenges, we propose a spectrum foundation model, termed SpectrumFM, which provides a new paradigm for spectrum cognition. An innovative spectrum encoder that exploits the convolutional neural networks and the multi-head self attention mechanisms is proposed to effectively capture both fine-grained local signal structures and high-level global dependencies in the spectrum data. To enhance its adaptability, two novel self-supervised learning tasks, namely masked reconstruction and next-slot signal prediction, are developed for pre-training SpectrumFM, enabling the model to learn rich and transferable representations. Furthermore, low-rank adaptation (LoRA) parameter-efficient fine-tuning is exploited to enable SpectrumFM to seamlessly adapt to various downstream spectrum cognition tasks, including spectrum sensing (SS), anomaly detection (AD), and wireless technology classification (WTC). Extensive experiments demonstrate the superiority of SpectrumFM over state-of-the-art methods. Specifically, it improves detection probability in the SS task by 30\% at -4 dB signal-to-noise ratio (SNR), boosts the area under the curve (AUC) in the AD task by over 10\%, and enhances WTC accuracy by 9.6\%. 
\footnote{The source code is available at \url{https://github.com/ChunyuLiu188/SpectrumFM.git}}
\end{abstract}

\begin{IEEEkeywords}
Spectrum foundation model, spectrum sensing, anomaly detection, wireless technology classification.
\end{IEEEkeywords}

\section{Introduction}
The radio frequency (RF) is subject to the increasingly severe spectrum scarcity problem due to the rapid proliferation of connected devices and the emergence of diverse wideband communication services~\cite{10901421}. This strain is further intensified by the massive connectivity of the Internet-of-Things (IoT) and the stringent performance requirements of the sixth-generation (6G) networks. In order to address this issue, spectrum cognition has emerged as a key enabler for intelligent spectrum management, supporting enhanced interference mitigation, and dynamic access in the complex and dynamic environments~\cite{meaning}.

Spectrum cognition encompasses a wide range of tasks, including spectrum sensing, anomaly detection, and wireless technology classification, etc~\cite{10980392}. Recently, significant success has been achieved in spectrum cognition by leveraging machine learning techniques. For example, Zhang \textit{et al.}~\cite{10496203} developed SSwsrNet that employed a semi-supervised learning technique based on MixMatch. The method made use of both labeled and unlabeled data, effectively overcoming the issue of limited labeled samples. As a result, it enhances classification accuracy in situations where obtaining labeled data is challenging. In \cite{10509639}, Zhang \textit{et al.} introduced a novel Transformer architecture designed to efficiently capture both intra-band spectrum characteristics and inter-band spectrum occupancy correlations within wideband signals. To tackle the difficulties associated with wideband spectrum sensing in scenarios where data is limited or cross-domain adaptation is required, Hao \textit{et al.}~\cite{10758375} proposed an innovative pre-training and fine-tuning strategy. By adopting transfer learning techniques, the method enhances model performance even when labeled data is scarce. Kang \textit{et al.}~\cite{9863875} developed an enhanced deep support vector data description (SVDD) aimed at extracting low-dimensional features from samples represented in the time-frequency domain to detect abnormal signals. The method not only delivered outstanding detection performance but also preserved real-time processing capabilities. The authors in \cite{10765508} introduced a new deep learning framework designed to perform both spectrum sensing and anomaly detection concurrently. The unified method delivered superior detection performance while ensuring robust real-time processing capabilities.

Although these methods have achieved promising results in various aspects of spectrum cognition, most current  methods are task-specific and designed to address only a single task and thus suffer from several fundamental limitations. Specifically, they often struggle with adapting to new or unseen environments, demonstrating limited generalization capability. Additionally, these methods typically rely on large-scale labeled datasets for effective training, which can be costly and time-consuming to obtain. Furthermore, their performance tends to degrade significantly in challenging scenarios such as low signal-to-noise ratio (SNR) conditions or under non-stationary signal dynamics.

Recent advances in foundation models have demonstrated a promising paradigm for building unified and generalizable frameworks to support diverse downstream tasks in remote sensing~\cite{10490262} and integrated sensing and communication (ISAC)~\cite{yang2025wirelessgptgenerativepretrainedmultitask}. These capabilities open up new possibilities beyond traditional supervised pipelines, enabling more scalable and robust solutions. However, the dynamic nature of the spectrum, with issues like interference and fading, complicates model generalization. Moreover, the wide range of frequency bands and unpredictable cognitive radio systems add to the complexity. Thus, applying foundation models to spectrum cognition is largely unexplored.

To this end, a novel spectrum foundation model, termed SpectrumFM, is proposed, which provides a new paradigm for spectrum cognition. In SpectrumFM, a novel spectrum encoder that exploits convolutional neural networks (CNNs) with multi-head self-attention (MHSA) is designed to effectively capture both fine-grained local signal structures and high-level global dependencies within the spectrum data. To facilitate pre-training on large-scale in-phase and quadrature (IQ) data, two novel self-supervised learning objectives, namely masked reconstruction and next-slot signal prediction, are proposed. These objectives enable the model to effectively learn robust and transferable representations. Furthermore, low-rank adaptation (LoRA) parameter-efficient fine-tuning is exploited to enable SpectrumFM to adapt to diverse downstream spectrum cognition tasks, including spectrum sensing (SS), anomaly detection (AD), and wireless technology classification (WTC). The fine-tuning can achieve superior performance while requiring the adjustment of only 2\% of its parameters. Extensive experiments demonstrate that SpectrumFM significantly outperforms state-of-the-art methods across multiple spectrum cognition tasks, validating its effectiveness and strong generalization ability. Specifically, it improves detection probability by 30\% in the SS task at a -4 dB SNR, boosts the area under the curve (AUC) by over 10\% in the AD task, and achieves a 6.8\% accuracy gain in the WTC task.

The remainder of this paper is organized as follows. Section~\ref{sec:model} provides a detailed description of the proposed model, including its spectrum encoder architecture, the two pre-training tasks, and the LoRA-based fine-tuning strategy. Section~\ref{sec:experiments} presents the experimental setup and the results obtained on downstream tasks. Finally, Section~\ref{sec:conclusion} concludes the paper and outlines directions for future work.

\section{Our Proposed Model}
\label{sec:model}
\begin{figure*}[htbp]
    \centering
    \includegraphics[width=0.9\textwidth]{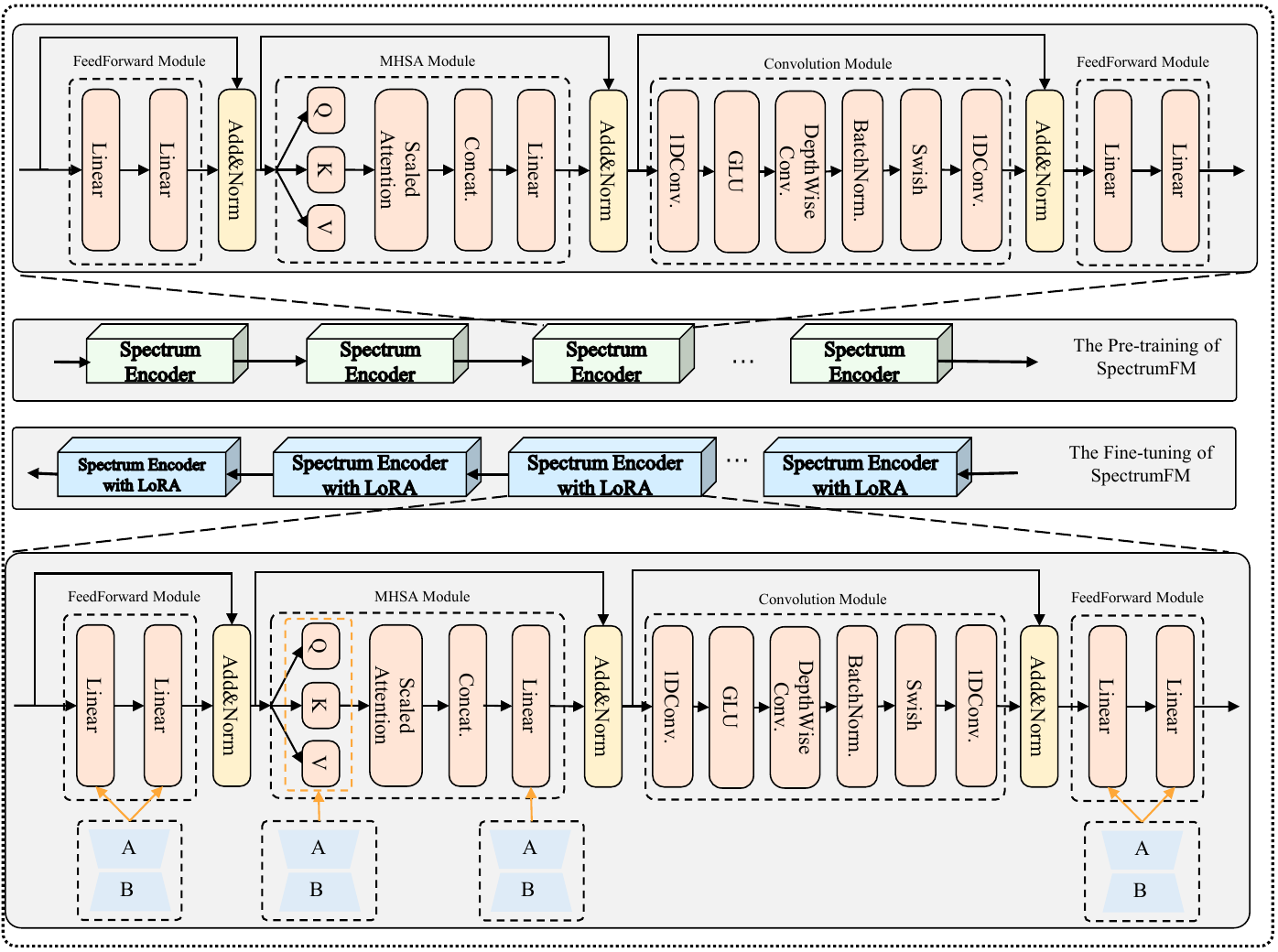}
    \caption{Overview of SpectrumFM, featuring a novel spectrum encoder that exploits CNNs with MHSA to
    jointly model fine-grained local signal structures and high-level global dependencies, a self-supervised pre-training to learn comprehensive and transferable representations, and LoRA fine-tuning for efficient adaptation to downstream spectrum cognition task, including SS, AD, and WTC.}
    \label{fig:model}
\end{figure*}

The framework of our proposed model is shown in Fig.~\ref{fig:model}. At the core of the framework is a novel spectrum encoder, which exploits CNNs with MHSA to effectively model fine-grained local signal structures and high-level global dependencies within the spectrum data. The framework is structured into two key stages, namely, the pre-training stage and the fine-tuning stage. During the pre-training stage, the encoder is trained on large-scale IQ data by leveraging two novel self-supervised learning objectives, namely masked reconstruction and next-slot signal prediction, enabling SpectrumFM to learn robust and generalizable spectrum representations. In the fine-tuning stage, LoRA is exploited to efficiently refine the model for various downstream spectrum cognition tasks, including SS, AD, and WTC, achieving high performance while requiring minimal parameter adjustments.
\subsection{Architecture of the Spectrum Encoder}
For each IQ signal $\mathbf{x}_{\text{iq}}$, the transformation process begins by converting the raw IQ data into its amplitude and phase (AP) components $\mathbf{x}_{\text{ap}}$. The amplitude, which is always positive and falls within the range $[0, +\infty]$, while the phase lies within the range $[-\pi, \pi]$. The transformation not only simplifies the normalization process but also ensures better consistency when handling signals from different sources. The normalization is then performed on each sample individually, scaling the values based on its own maximum and minimum, thereby mapping them to a standardized range, given as 
\begin{equation}
    \mathbf{x}^{\text{norm}}_{\text{ap}}=\frac{x_{\text{ap}}-\min(x_{\text{ap}})}{\max(x_{\text{ap}})-\min(x_{\text{ap}})},
    \label{eq:1}
\end{equation}
where $\min(\cdot)$ and $\max(\cdot)$ are the functions to obtain the minimum and maximum values of the sample $x_{\text{ap}}$, respectively. After obtaining the normalized signal, a 1D convolution layer is leveraged to project the normalized signal into a higher-dimensional space, corresponding to the hidden layer dimension $d$, given as 
\begin{equation}
    \mathbf{x}_{\text{proj}} = \text{Conv1D}(\mathbf{x}_{\text{ap}}^{\text{norm}}, \mathbf{W}_{\text{proj}}),
    \label{eq:2}
\end{equation}
where $\mathbf{W}_{\text{proj}}$ is the kernel martix.
To enable the model to understand the sequential order of the signal, positional encoding is added to $\mathbf{x}_{\text{proj}}$. The positional encoding for the position \(p\) in the sequence and the dimension \(i\) of the hidden space is given as
\begin{subequations}
    \begin{align}
        \text{PE}_{p, 2i} &= \sin\left(\frac{p}{10000^{2i/d}}\right),\\
        \text{PE}_{p, 2i+1} &= \cos\left(\frac{p}{10000^{2i/d}}\right),
    \end{align}
\end{subequations}
where \(p\) is the position index, and \(i\) is the index of the hidden dimension.

The processed signal \(\mathbf{x}_{\text{position}}\) is then passed into the proposed spectrum encoder for feature extraction. The encoder follows a well-defined pipeline consisting of several key modules, namely, an initial feedforward module, a MHSA module, a convolution module, and a final feedforward module. Each of the modules is connected via residual connections, followed by normalization, to ensure stability during training.

The process begins with the input sequence undergoing initial transformation through a feedforward module, which enhances the feature representation by applying non-linear transformations. Specifically, the feedforward module involves two successive linear transformations separated by a non-linear activation function. The output of the module is given as 
\begin{equation}
    \mathbf{x}_{\text{ffn}} = \text{RN}\left(\text{GELU}(\mathbf{x}_{\text{position}} \mathbf{W}_1 + \mathbf{b}_1)\mathbf{W}_2 + \mathbf{b}_2\right),
\end{equation}
where $\text{RN}$ denotes the residual connection and normalization operation, $\text{GELU}$ denotes the Gaussian error linear unit (GELU) activation function, and $\mathbf{W}_1, \mathbf{W}_2, \mathbf{b}_1$, and $\mathbf{b}_2$ are the weight matrices and bias vectors.

The enriched feature representation $\mathbf{x}_{\text{ffn}}$ is then processed by the MHSA module, enabling the model to capture long-range dependencies across the sequence. For each attention head $h$, the input $\mathbf{x}_{\text{ffn}}$ is projected into query ($\mathbf{Q}$), key ($\mathbf{K}$), and value ($\mathbf{V}$) matrices by using learnable weight matrices, given as 
\begin{subequations}
    \begin{align}
        \mathbf{Q}_h &= \mathbf{x}_\text{ffn} \mathbf{W}_Q^h, \\
        \mathbf{K}_h &= \mathbf{x}_\text{ffn} \mathbf{W}_K^h, \\
        \mathbf{V}_h &= \mathbf{x}_\text{ffn} \mathbf{W}_V^h,
    \end{align}
\end{subequations}
where $\mathbf{W}_Q^h$, $\mathbf{W}_K^h $ and $\mathbf{W}_V^h$ are the learnable weight matrices.
Each head computes its output through a scaled dot-product attention mechanism, which is given as 
\begin{equation}
    \mathbf{x}_h = \text{softmax}\left(\frac{\mathbf{Q}_h \mathbf{K}_h^T}{\sqrt{d_h}}\right) \mathbf{V}_h,
\end{equation}
where $d_h = \frac{d}{H}$ denoting the dimension of the head, and $H$ is the number of attention heads.
The outputs from all attention heads are concatenated and then linearly transformed to produce the final output of the MHSA module, given as 
\begin{equation}
    \mathbf{x}_{\text{attention}} = \text{RN}\left(\text{concat}(\mathbf{x}_1, \dots, \mathbf{x}_H) \mathbf{W}_o\right),
\end{equation}
where concat is the concatenation operator, and $\mathbf{W}^o$ is the weight matrix.

Subsequently, the output from the MHSA mechanism is processed through a series of convolutional layers. First, a 1D convolution with a kernel size of 1 is applied, followed by a depthwise separable convolution with a kernel size of 3. Finally, another 1D convolution is applied to capture fine-grained features. Mathematically, the sequence can be expressed as
\begin{equation}
    \mathbf{x}_{\text{conv1}} = \text{Conv1D}(\mathbf{x}_{\text{attention}}, \mathbf{W}_{\text{conv1}}),
\end{equation}
where $\mathbf{W}_{\text{conv1}}$ is the trainable weight matrix for the first 1D convolution.
Next, the depthwise separable convolution with a kernel size of 3 is applied as
\begin{equation}
    \mathbf{x}_{\text{conv2}} = \text{DepthwiseConv1D}(\mathbf{x}_{\text{conv1}}, \mathbf{W}_{\text{conv2}}),
\end{equation}
where $\mathbf{W}_{\text{conv2}}$ is the trainable weight matrix for the depthwise separable convolution.
Finally, the last 1D convolution layer is applied as
\begin{equation}
    \mathbf{x}_{\text{conv3}} = \text{Conv1D}(\mathbf{x}_{\text{conv2}}, \mathbf{W}_{\text{conv3}}),
\end{equation}
where $\mathbf{W}_{\text{conv3}}$ is the weight matrix for the final 1D convolution.

After the MHSA and convolution modules, the resulting feature $\mathbf{x}_{\text{conv3}}$ is further refined through a final feedforward module. The step integrates the benefits of both MHSA and convolution modules, given as 
\begin{equation}
    \mathbf{x}_{\text{ffn2}} = \text{RN}(\text{GELU}(\mathbf{x}_{\text{conv3}} \mathbf{W}_3 + \mathbf{b}_3)\mathbf{W}_4 + \mathbf{b}_4),
\end{equation}
where $\mathbf{W}_3$ and $\mathbf{W}_4$ are the trainable weight matrices, $\mathbf{b}_3$ and $\mathbf{b}_4$ are the bias vectors.

To enhance model expressiveness, the spectrum encoder block is stacked $L$ times to form the complete encoder. The final output of the $L$-th encoder block is denoted as $\mathbf{x}_{\text{hidden}}^{(L)}$.
\subsection{Pre-Training Tasks}
In this section, we introduce the pre-training tasks exploited to pre-train the spectrum encoder. The mask reconstruction is well-suited for spectrum data as it enables the model to learn robust signal representations by recovering degraded or missing segments, reflecting real-world signal impairments. The next-slot signal prediction complements this by capturing the dynamic nature of spectrum usage, allowing the model to anticipate future changes and adapt effectively to evolving environments.
\subsubsection{Masked Reconstruction Task}
The masked reconstruction task is designed to train the encoder to reconstruct the input signals with masked symbols. 
Given the normalized sequence $\mathbf{x}_{\text{ap}}^{\text{norm}}$, a binary mask vector $\mathbf{m} \sim \text{Bernoulli}(1 - r)$ is sampled, where $\{m_1, m_2, \ldots, m_N\}$ represents the individual elements. Specifically, $m_p=0$ indicates that the $p$-th element is masked, whereas $m_p=1$ signifies it remains unmasked. The masked input is defined as
\begin{equation}
    \mathbf{x}_{\text{masked}}[p] = 
    \begin{cases}
    0, & \text{if } m_p = 0, \\
    \mathbf{x}_{\text{ap}}^{\text{norm}}[p], & \text{if } m_p = 1.
    \end{cases}
\end{equation}
The masked input is first processed by the encoder, generating the hidden representation $\mathbf{x}_{\text{hidden}}^{(L)}$. A lightweight decoder, consisting of two linear layers, is then employed to reconstruct the original values at the masked positions. The reconstruction loss is computed by using the mean squared error (MSE) loss, formulated as
\begin{equation}
    \mathcal{L}_{\text{recon}} = \frac{1}{\sum_{p=1}^N (1 - m_p)} \sum_{p=1}^N (1 - m_p) \cdot \left\| \hat{\mathbf{x}}_{\text{norm}}^{\text{ap}}[p] - \mathbf{x}_{\text{norm}}^{\text{ap}}[p] \right\|_2^2,
\end{equation}
where $\hat{\mathbf{x}}_{\text{norm}}^{\text{ap}}$ is the predicted values, $N$ is the number of signal symbols, and $\left\| \cdot \right\|$ is the euclidean norm.
\subsubsection{Next-Slot Signal Prediction Task}
In addition to the masked reconstruction task, the encoder is also trained to predict the next signal symbol based on the observed sequence. Specifically, given the first $N-1$ points, the model is trained to predict the $N$-th point.
The prediction loss is computed by using the MSE loss between the predicted and true final point, given as 
\begin{equation}
    \mathcal{L}_{\text{pred}} = \left\| \hat{\mathbf{x}}[N] - \mathbf{x}[N] \right\|_2^2,
\end{equation}
where $\hat{\mathbf{x}}_{N}$ denotes the predicted final point.
\subsection{LoRA-Based Fine-Tuning for Downstream Spectrum Cognition Tasks}
To efficiently adapt the pre-trained SpectrumFM for specific downstream spectrum cognition tasks while maintaining computational efficiency, the LoRA parameter-efficient fine-tuning technique is exploited. 
Specifically, given a weight matrix $\mathbf{W} \in \mathbb{R}^{d \times d}$ within the encoder layers, LoRA decomposes the adaptation into two low-rank matrices, $\mathbf{A} \in \mathbb{R}^{d \times a}$ and $\mathbf{B} \in \mathbb{R}^{a \times d}$, where $a \ll d$. The adapted weight matrix is then expressed as
\begin{equation} \mathbf{W}_{\text{LoRA}} = \mathbf{W} + \alpha \mathbf{A} \mathbf{B}, \end{equation} 
where $\alpha$ serves as a scaling factor to regulate the impact of the low-rank update. The pre-trained weight matrix $\mathbf{W}$ remains frozen, while the low-rank matrices $\mathbf{A}$ and $\mathbf{B}$ introduce minimal additional parameters. In terms of computational complexity, LoRA incurs a parameter overhead of $\mathcal{O}(ad)$, which is significantly lower than the full fine-tuning method that requires $\mathcal{O}(d^2)$ parameters.
On top of the encoders, task-specific heads are designed. These task heads primarily utilize linear layers for final decision-making. However, for WTC and SS tasks, a gated recurrent unit (GRU) is first introduced to aggregate hidden representations before passing the processed features to the linear layers, ensuring effective feature integration for classification.

\section{Experiments}
\label{sec:experiments}
\subsection{Experimental Seetings}
The key hyperparameters for SpectrumFM are as follows. The mask ratio $r$ is set to $15\%$, with a hidden dimension of $d = 256$ and a feedforward dimension of 512. The encoder consists of $L = 16$ layers, while the number of symbols $N$ is set to 128. A dropout ratio of 0.1 is applied, and the scaling factor $\alpha$ is set to 16, with a low-rank value of $a = 8$.
For pre-training, the model undergoes 10 epochs using a batch size of 256 and a learning rate of 0.001, optimized via AdamW, with early stopping employed to mitigate overfitting. During fine-tuning, the learning rate is set to 0.001, the batch size to 256.
\subsection{Spectrum Sensing Task}
The SS task involves detecting primary users within a frequency band to enable dynamic spectrum access in cognitive radio networks. The RML2018.01A~\footnote{\url{https://www.deepsig.ai/datasets/}} dataset, which includes various modulation schemes, is augmented with pure noise to represent unoccupied spectrum regions. It allows evaluation of spectrum sensing algorithms ability to distinguish between occupied and unoccupied frequencies. To evaluate the few-shot learning capabilities of the proposed model, a total of 6,000 samples are selected for the training process. Baseline models include GRU networks, which capture temporal dependencies in IQ data, and residual networks (ResNet), which utilize residual connections to extract spatial features.
\subsubsection{Performance Comparison}   
\begin{figure}[t]
    \centering
    \includegraphics[width=0.6\linewidth]{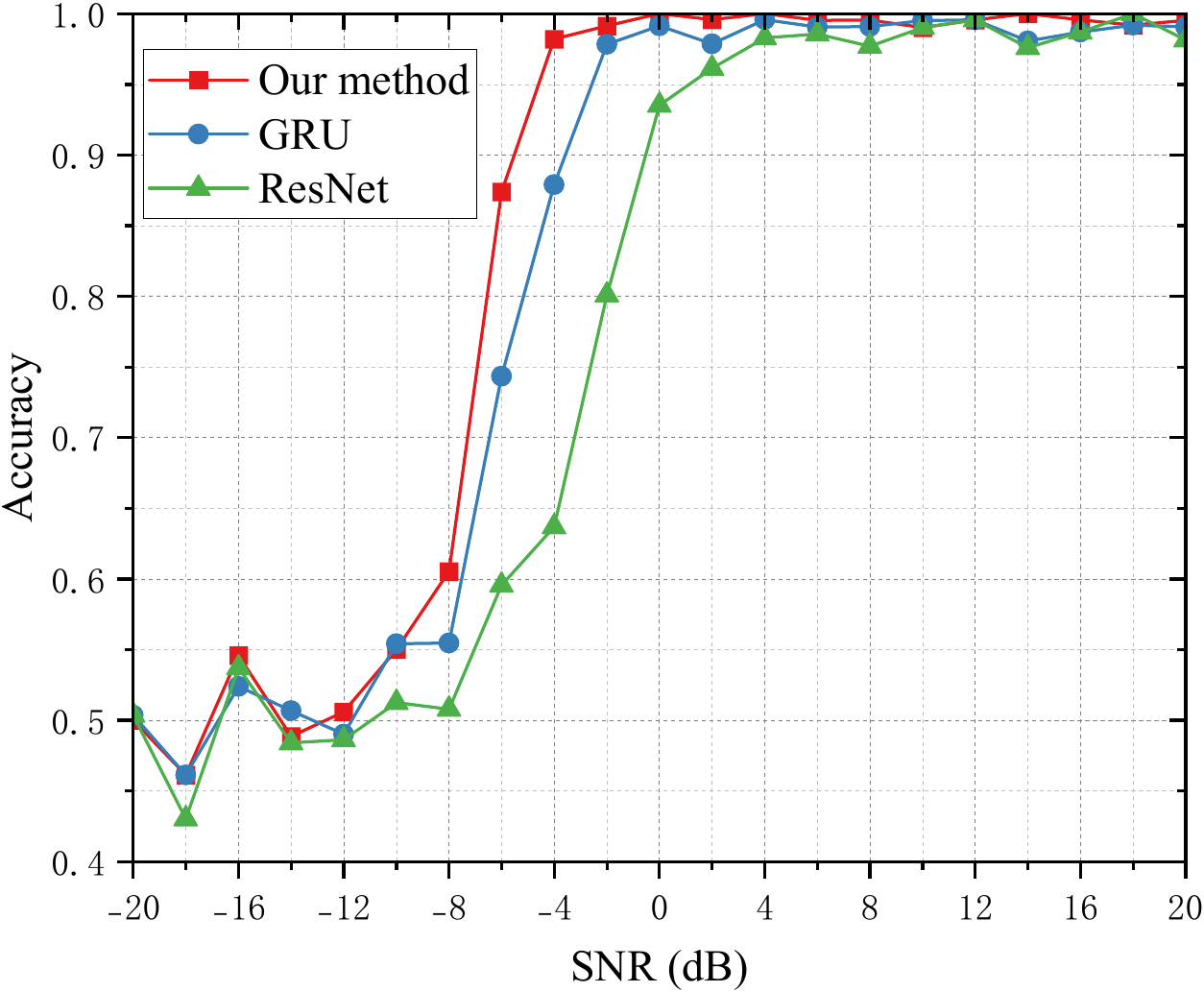}
    \caption{Sensing accuracy of our model and baseline models at
    various SNR levels in the SS task.}
    \label{fig:sensing_snr}
\end{figure} 
\begin{figure}[t]
    \centering
    \includegraphics[width=0.6\linewidth]{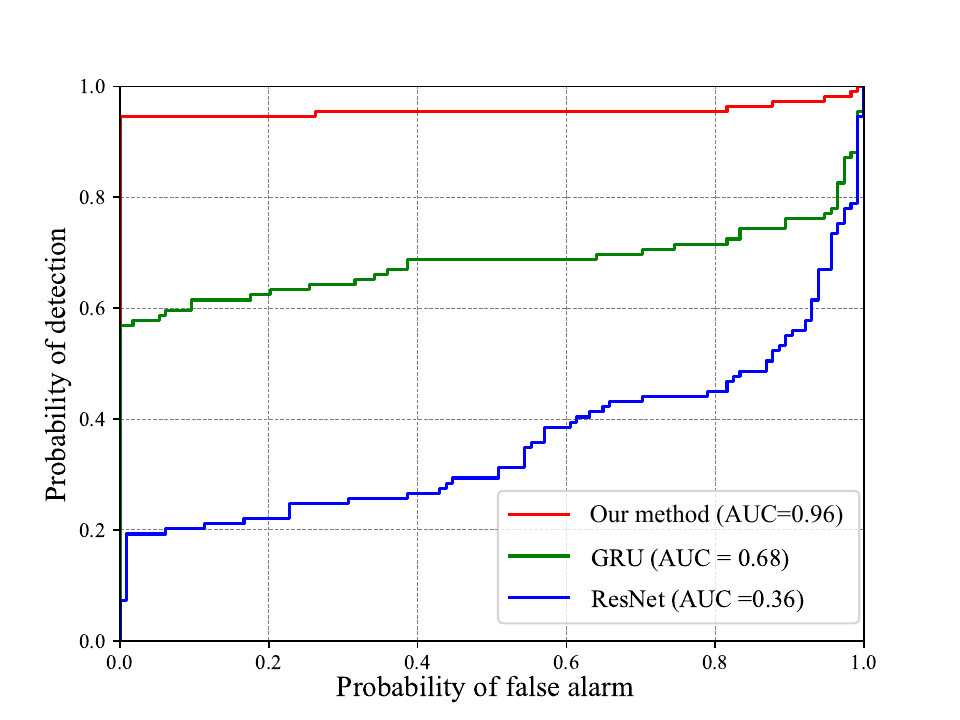}
    \caption{ROC curve of our model and baseline models at -4dB SNR level in the SS task.}
    \label{fig:sensing_ROC}
\end{figure} 
The performance of SS models across various SNR levels is shown in Fig.~\ref{fig:sensing_snr}. Our proposed model outperforms GRU and ResNet-based models in low SNR environments, demonstrating enhanced accuracy and reliability in challenging sensing environments. The ROC curves for different models at a SNR of -4 dB are shown in Fig.~\ref{fig:sensing_ROC}. It illustrates that our proposed model achieves a superior balance between detection probability and false alarm rate, showing higher accuracy in identifying true positives while minimizing false positives compared to the GRU and ResNet models. It indicates that even under challenging signal conditions, our model maintains robust spectrum sensing capabilities, offering enhanced reliability and performance over alternative methods. 
\subsection{Anomaly Detection Task}
The AD task aims to identify disruptions in wireless communication systems by distinguishing abnormal patterns from normal ones. Data collection is performed over a two-day period using an over-the-air (OTA) platform. The setup includes a ceyear 1435B-V RF signal generator transmitting QPSK-modulated primary user (PU) signals at a 2.9 GHz carrier frequency and a bandwidth of 20 MHz. To simulate challenging interference, aliased-signals with bandwidths of 10 MHz and 30 MHz are introduced, causing aliasing effects within the frequency range of 2.885 GHz to 2.915 GHz. The signals are also QPSK-modulated to increase the complexity of detection. A SAM 60 MK2 receiver captures the resultant signals for analysis. For the AD baseline model, an adversarial autoencoder (AAE)~\cite{10035489} is utilized, which combines an autoencoder for input reconstruction and a discriminator to enforce a structured latent space distribution, thereby enhancing the detection of anomalous patterns.
\subsubsection{Performance Comparison}
\begin{figure}[t]
    \centering
    \includegraphics[width=0.6\linewidth]{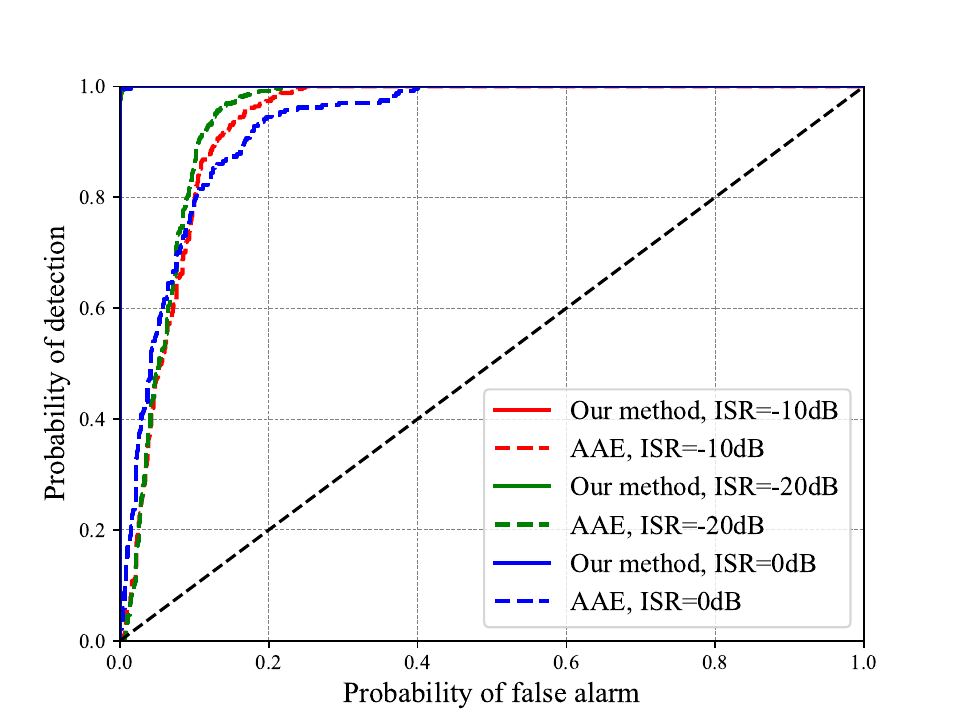}
    \caption{ROC curves for 10M aliased-signal interference.}
    \label{fig:ROC_curve_10M}
\end{figure}
\begin{figure}[t]
    \centering
    \includegraphics[width=0.6\linewidth]{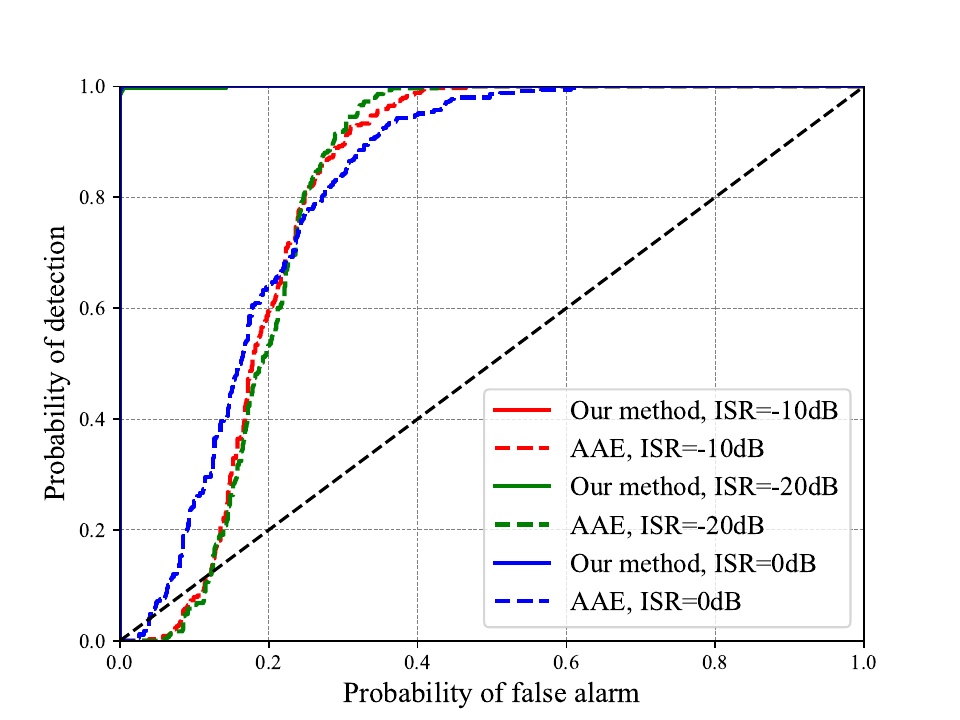}
    \caption{ROC curves for 30M aliased-signal interference.}
    \label{fig:ROC_curve_30M}
\end{figure}  
The ROC curves in Fig.~\ref{fig:ROC_curve_10M} and \ref{fig:ROC_curve_30M} illustrate the performance of our proposed model compared to the AAE baseline under different SNR levels for 10 MHz and 30 MHz aliased-signal interference, respectively.
For the 10 MHz aliased-signal interference, our model consistently outperforms the AAE across all SNR levels. Specifically, at SNRs of -10 dB, -20 dB, and 0 dB, our model achieves higher probabilities of detection with lower probabilities of false alarm, indicating superior robustness and accuracy.
For the 30 MHz aliased-signal interference, our model demonstrates better performance than the AAE. At all tested SNR levels, our model shows a higher probability of detection while maintaining a lower probability of false alarm, highlighting its effectiveness in handling more complex interference scenarios.
Overall, these results indicate that our proposed model is more effective in detecting abnormal signals under various SNR conditions.
\subsection{Wireless Technology Classification Task}
In the WTC task, the goal is to accurately identify various wireless communication technologies across different environments. 20,000 samples from the TechRec~\footnote{\url{https://ieee-dataport.org/documents/iq-signals-captured-lte-wifi-and-dvb-t}} dataset are utilized for training. The baseline models include AMC\_Net~\cite{10097070}, MSNet~\cite{9463441}, CGDNN~\cite{10261289} and MCNet~\cite{9915584}.
\subsubsection{Performance Comparision}
\begin{table}[t]
    \centering
    \caption{Comparison of Precision, Recall And F1-score in the WTC Task.}
    \label{tab:comparison}
    \begin{tabular}{c|c|c|c}
    \hline
    Models  & Precision & Recall & F1-score \\  \hline
    AMC\_Net & 0.6577 & 0.6441 & 0.6368 \\ \hline
    MCNet & 0.7131 & 0.7308 & 0.7114 \\ \hline
    CGDNN & \underline{0.7827} & 0.7448 & 0.7338 \\ \hline
    MSNet & 0.7597 & \underline{0.7509} & \underline{0.7480} \\ \hline
    Our method & \textbf{0.8218} & \textbf{0.8227} & \textbf{0.8216} \\ \hline
    \end{tabular}\end{table}
\begin{figure}[t]
    \centering
    \includegraphics[width=0.6\linewidth]{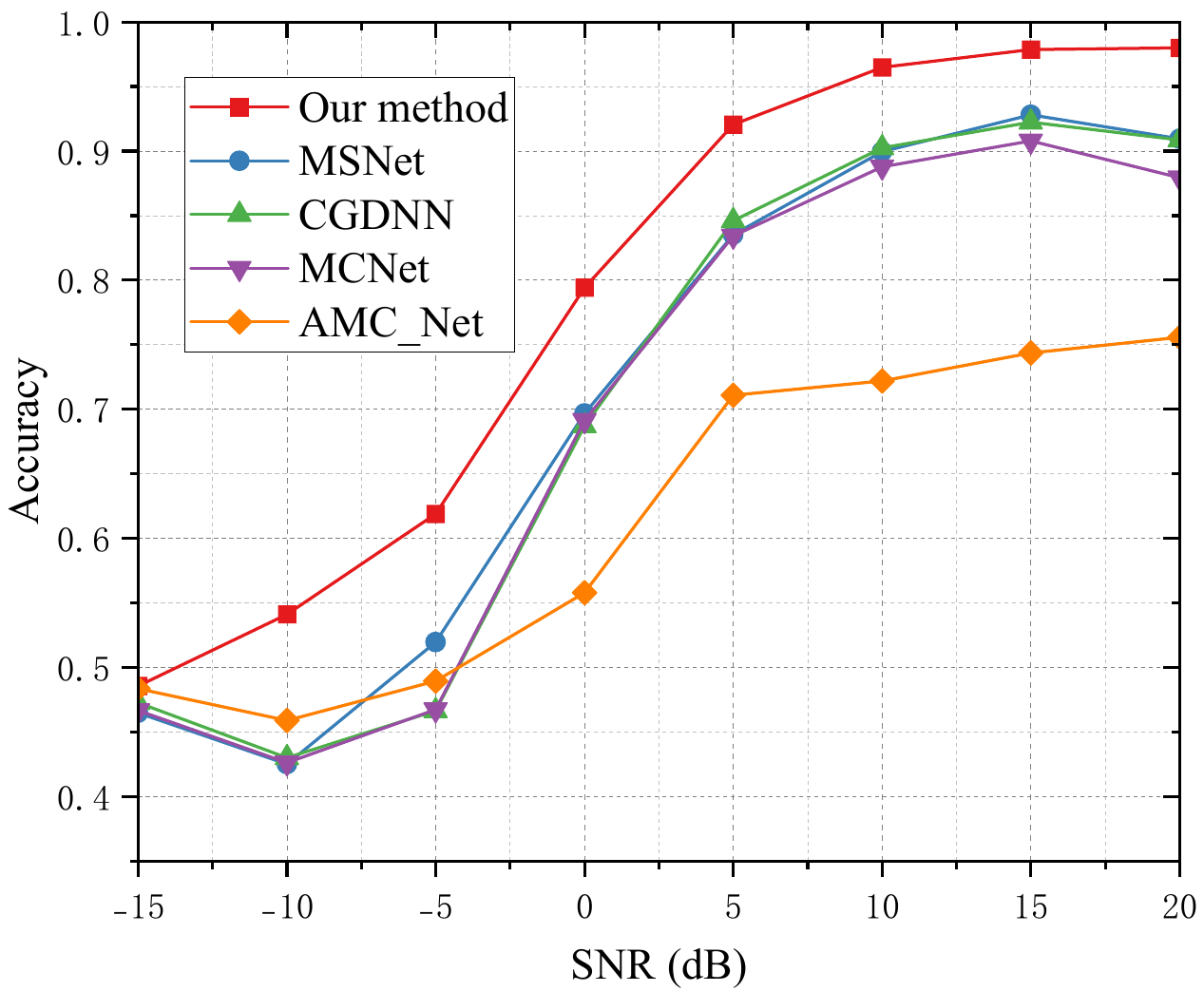}
    \caption{The accuracy of our model and baseline models at
    various SNR levels in the WTC task.}
    \label{fig:wtc}
\end{figure}  
TABLE~\ref{tab:comparison} presents the precision, recall, and F1-score of our model compared to the baseline models. The results demonstrate that our model consistently outperforms the baselines across all evaluation metrics, highlighting its superior ability to accurately identify different wireless communication technologies.
Fig.~\ref{fig:wtc} illustrates the accuracy of our model and baseline models across various SNR levels. Our model consistently outperforms the baselines, achieving higher accuracy across all SNR ranges. It indicates that our model is more robust and effective in accurately identifying wireless communication technologies under different signal conditions.
\section{Conclusion}
\label{sec:conclusion}
In this paper, a foundation model, termed SpectrumFM, was proposed, which provides a new paradigm for spectrum cognition.
A novel spectrum encoder exploiting CNNs with MHSA mechanisms was proposed to effectively extract both fine-grained local signal structures and high-level global dependencies.
To pre-train SpectrumFM, two novel self-supervised pre-training objectives, namely masked reconstruction and next-slot signal prediction, were introduced to enable SpectrumFM to learn comprehensive and transferable representations. Furthermore, the LoRA parameter-efficient fine-tuning technique was exploited, enabling SpectrumFM to efficiently adapt to various spectrum cognition tasks, including SS, AD, and WTC, while adjusting only 2\% of the total parameters.
Extensive experiments demonstrated the strong generalization ability and superior performance of SpectrumFM across various spectrum cognition tasks. Specifically, SpectrumFM improves detection probability by 30\% in the SS task at a -4 dB SNR, boosts AUC by over 10\% in AD, and surpasses the state-of-the-art by 6.8\% in accuracy for WTC. For the future work, we believe that SpectrumFM has the potential to support dynamic spectrum access, efficient spectrum resource allocation, and secure spectrum sharing.

\bibliographystyle{IEEEtran} 
\bibliography{ref}

\end{document}